# On IEEE 802.11: Wireless LAN Technology

**Sourangsu Banerji[1], Rahul Singha Chowdhury[2]**
[1, 2] Department of Electronics & Communication Engineering,
RCC-Institute of Information Technology, India

**ABSTRACT: Network technologies are traditionally based on wireline solutions. But the introduction of the IEEE 802.11 standards have made a huge impact on the market such that laptops, PCs, printers, cellphones, and VoIP phones, MP3 players in our homes, in offices and even in public areas have incorporated the wireless LAN technology. Wireless broadband technologies nowadays provide unlimited broadband access to users which were previously offered only to wireline users. In this paper, we review and summarize one of the emerging wireless broadband technology i.e. IEEE 802.11,which is a set of physical layer standard for implementing wireless local area network computer communication in the 2.4,3.6,5 and 60GHz frequency band. They fix technology issues or add functionality which is expected to be required by future applications. Though some of the earlier versions of these technologies are obsolete (such as HiperLAN) now but still we have included them in this review for the sake of completeness.**

**Keywords:** Wireless Communications, IEEE 802.11, HiperLAN, WLAN, Wi-fi.

## 1. Introduction

The wireless broadband technologies were developed with the aim of providing services comparable to those provided to the wireline networks. Cellular networks now provide support for high bandwidth data transfer for numerous mobile users simultaneously. In addition to this, they also provide mobility support for voice communication. Wireless data networks can be divided into several types depending on their area of coverage. They are:
WLAN: Wireless Local Area network, in area with a cell radius up to hundred meters, mainly in home and office environments.

WMAN: Wireless Metropolitan Area Network; generally cover wider areas as large as entire cities. WWAN: Wireless Wide Area Network with a cell radius about 50 km, cover areas larger than a city.

However out of all of these standards, WLAN and recent developments in WLAN technology would be our main area of study in this paper. The IEEE 802.11 is the most widely deployed WLAN technology as of today. Another well known is the HiperLAN standard by ETSI. Both these technologies are united under the Wireless Fidelity (Wi-fi) alliance. In literature though, IEEE802.11 and Wi-fi is used interchangeably and we will also follow the same convention in this paper. A typical WLAN network consists of an Access Point (AP) in the middle/centre and a number of stations (STAs) are connected to this central Access Point (AP).Now, there are basically two modes in which communication can take place.

In the centralized mode of communication, communication to/from a STA is always carried over the APs. There is also a decentralized mode in which communication between two STAs can take place directly without the requirement of an AP in an ad hoc fashion. WLAN networks provide coverage up to an area of 50-100 meters. Initially, Wi-fi provided an aggregate throughput of 11Mbps, but recent developments have increased the throughput to about 54 Mbps. As a result of its high market penetration, several amendments to the basic IEEE 802 standard have been developed or are currently under development.

In this paper, we overview the IEEE 802.11 standard and address the technical context of its extensions. In section 2, we briefly discuss the history behind the development of the standard. Section 3 deals with the features of the IEEE 802 family which have already been implemented. In the following Section 4, we look at other standards besides IEEE 802.11a; IEEE 802.11b/g. In section 5, some of the upcoming standards are discussed and some open issues with the IEEE 802.11 standard is given in section 6. The HiperLAN and its standards have been talked about in section 7 and 8. In Section 9 we state the reasons of HiperLANs' failure over IEEE 802.11. Lastly we conclude our paper in section 10.

## 2. Development of IEEE 802.11

The Physical layer (PHY) and medium access control (MAC) layer were mainly targeted by the IEEE 802 project. When the idea of wireless local area network (WLAN) was first conceived, it was just thought of another PHY of one of the available standards. The first candidate which was considered for this was IEEE's most prominent standard 802.3.

However later findings showed that the radio medium behaved quite different than the conventional well behaved wire. As there was attenuation even over short distances, collisions could not be detected. Hence, 802.3's carrier sense multiple access with collision detection (CSMA/CD) could not be applied.

The next candidate standard considered was 802.4. At that point of time, its coordinated medium access i.e. the token bus concept was believed to be superior to 802.3's contention-based scheme. Hence, WLAN began as 802.4L. Later in 1990 it became obvious that token handling in radio networks was rather difficult. The standardization body realized the need of a wireless communication standard that would have its own very unique MAC. Finally, on March 21, 1991, the project 802.11 was approved (fig. 1).

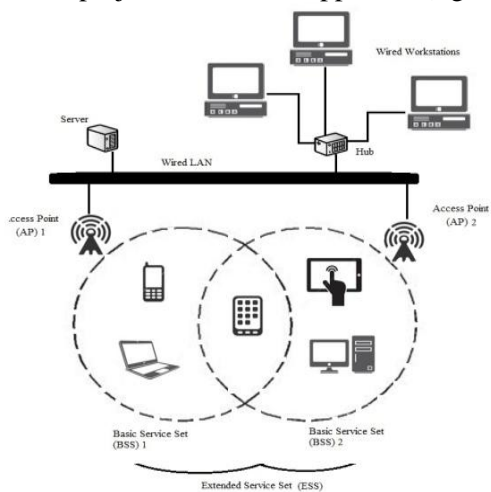

**Figure 1** WLAN Network Architecture

## 3. IEEE 802.11 family

The most widely deployed 802.11 standard has a lot of extension and many more are currently under development. First introduced in 1999,the IEEE 802.11 standards were primarily developed keeping in mind the home and the office environment for wireless local area connectivity. The Initial standards gave a maximum data rate of 2Mbps per AP which increased to 11 Mbps per AP with the deployment of IEEE 802.11b [2].Newer extensions like IEEE 802.11g and IEEE 802.11a provided maximum data rate of 54Mbps per AP using various methods to boost up the maximum data rates [3-5]. WLAN devices based on IEEE 802.11g currently offer data rate 100-125Mbps [4].

Similarly, a relatively newer IEEE 802.11n gives a maximum data rate of about 540Mbps [25].Furthermore, in addition to these, several other standards were deployed which solved many QoS and security issues related with the earlier standards. Additional mechanisms were introduced to remedy QoS support and security problems in IEEE 802.11e [12] and IEEE 802.11i.The IEEE 802.11n standard which we earlier talked about also introduced MAC enhancements to overcome MAC layer limitations in the current standards [28]. The IEEE 802.11s standard added mesh topology support to the IEEE 802.11 [34]. The IEEE 802.11u improved internetworking with external non-802.11 networks. The IEEE 802.11w was an added onto 802.11i covering management frame security.

The IEEE 802.11ad standard adds a "fast session transfer" feature, enabling the wireless devices to seamlessly make transition between the legacy 2.4 GHz and 5 GHz bands and the 60 GHz frequency band [41]. The IEEE 802.11ac standard, still under development is expected to provide a multi-station WLAN throughput of at least 1 Gbps and a single link throughput of at least 500 Mbps [45].

### 3.1. Physical (PHY) Layer

The IEEE 802.11 uses variety of PHY layers with the aim of increasing the aggregate throughput of the network. IEEE 802.11 standard includes three PHY layers namely:
1. FHSS (Frequency Hopping Spread Spectrum)

**Table 1**
OFDM PHY layer modulation techniques

| Data Rate (Mbps) | Modulation | Coding rate | Coded bits/sub carrier | Code bits/OFDM symbol | Data bits/OFDM symbol |
|---|---|---|---|---|---|
| 6 | BPSK | 1/2 | 1 | 48 | 24 |
| 9 | BPSK | 3/4 | 1 | 48 | 36 |
| 12 | QPSK | 1/2 | 2 | 96 | 48 |
| 18 | QPSK | 3/4 | 2 | 96 | 72 |
| 24 | 16-QAM | 1/2 | 4 | 192 | 96 |
| 36 | 16-QAM | 3/4 | 4 | 192 | 144 |
| 48 | 64-QAM | 2/3 | 6 | 288 | 192 |
| 54 | 64-QAM | 3/4 | 6 | 288 | 216 |

2. DSSS (Direct Sequence Spread Spectrum)
3. IR (Infrared)

In addition to these 802.11b uses a new PHY layer, High Rate DSSS [1].IEEE 802.11a and 802.11g are based on OFDM (Orthogonal Frequency Division Multiplexing); that greatly increases the overall throughput of the AP [2-4]. Various OFDM modulation techniques are summarized in Table 1. 802.11n also uses OFDM modulation technique but coupled with a MIMO (Multi Input Multi Output) mechanism [28].The frequency band of operation of most of the extensions of IEEE 802.11 is 2.4 GHz with 14 distinct channels. The availability of these channels varies from one country to another. Out of these the last channel was especially designed for Japan which was the main feature incorporated in IEEE 802.11j extension.

The IEEE 802.11a extension employs a number of channels ranging from 36-161 depending on the frequency band (5.15-5.825 GHz) although it works with a fixed channel centre frequency of 5 GHz. There are 12 non overlapping channels in the frequency band for the IEEE standard in the U.S. and 19 non-overlapping channels in Europe [3]. In contrast, there are only 3 out 14 non-overlapping in case of 802.11b [2]. IEEE 802.11n uses overlapping channels with channel bandwidth 20 and 40MHz [19]. The 20MHz channel bandwidth is incorporated in every 802.11n device, the 40MHz channel is optional.

Peer to Peer (P2P) WLAN links can be established with the help of directional antennas for a few km ranges. A typical WLAN Access Point (AP) uses omnidirectional antennas with a range of 30-50m (indoors) and 100m (outdoors).This range is greatly affected by the obstacles between the AP and the STA.IEEE 802.11a suffer from increased range and attenuation compared to IEEE 802.11b/g networks, because it operates on the higher frequency range of 5MHz.Use of sectored antennas instead of omnidirectional antennas increases the aggregate WLAN data rate in a given area to 2-3 times [6].

## 3.2. Medium Access Control (MAC) Layer

IEEE 802.11 uses a contention based scheme known as Distributed Coordinated Function (DCF). In this method the STA linked with the AP scans the air interface for channel availability. If the interface is idle, the STA sends it data to the destination through the AP. If however the air interface is busy or more than one STA tries to access the same AP; a collision occurs. The IEEE 802.11 uses a Carrier Sense Multiple Access/Collision Avoidance (CSMA/CA) to avoid the collisions. IEEE 802.11 uses another MAC technique known as Point Coordination Function (PCF) [18]. This mechanism is divided in to two parts. In the first part, the AP scans all its STA in a round robin fashion and checks to see if any of the STAs has any packets to send. If any of the STAs is not polled during the current period, it will be queued up for polling during the next polling period. The scanned part uses the contention based scheme and it same as DCF.

Moreover, due to polling mechanism in PCF the aggregate throughput of an IEEE 802.11 network decreases. DCF is the default MAC technique used in the IEEE 802.11 standard. While the standard includes both the MAC techniques, PCF is included in the Wi-fi alliance standard and hence not quite as popular as DCF [17]. In both the MAC techniques an automatic response request mechanism is used in this method. Any device in the network receiving data will send an acknowledgement signal (ACK) back to the sender. In case the receiver receives a corrupt data packet, it issues a NAK (Negative Acknowledgement) and the sender resends the data packet. There is a round trip delay as the sender has to wait for the ACK to transmit the next data packet in the queue.

### 3.3. Request to Send/Clear to Send (RTS/CTS)

In the contention based scheme called DCF if more than two STAs simultaneously try to access the air interface, a collision occurs. To avoid such collision CSMA/CA may result in incorrect medium information. This is called Hidden Node Problem in which collision in the some part of the network cannot be detected [15]. If any two STAs cannot directly communicate, the AP invokes a RTS/CTS mechanism. For each transmission, the source STA issues a RTS message. The destination STA replies to this by sending a CTS message. Upon receiving the CTS message, the source STA starts its data transmission. The medium is assumed to be in use given in the message when they receive RTS and/or CTS message. In PCF using RTS/CTS reduces the network throughput.

### 3.4. Authentication & Encryption

Security is also handled in the MAC layer. To avoid unauthorized access from other STAs, several encryption methods have been used. One of earlier encryption mechanism was Wired Equivalent Piracy (WEP). But the encryption method had security vulnerabilities and the Wi-fi Alliance developed another encryption technique named Wi-Fi protection Access (WPA).
The IEEE 802.11i standard incorporated an enhanced version of WPA (WPA2) [20]. IEEE 802.11i also addressed security issues associated with authentication methods like open standard and shared key authentication and incorporated IEEE 802.1X authentication method which is now used in all the later versions of IEEE 802 family standards. In this method, users can authenticate their identities by a RADIUS or diameter server.

### 3.4.1. Management Frame

The current 802.11 standards define "frame" types for use in management and control of wireless links. The TGw implemented the IEEE 802.11w standard to implement the Protected Management Frames. The TGw is still working on improving the IEEE 802.11 MAC layer. Security can be enhanced by providing data confidentiality of management frames. These extensions will have interactions with IEEE 802.11r as well as IEEE 802.11u

### 3.5. Operating Modes

The IEEE 802.11 supports two operating modes. They are namely
1. Infrastructure operating mode and
2. Independent operating mode.

In the infrastructure operating mode, the STAs communicate with each other through the Access Point. In this scheme, an STA needs to be connected to an Access Point in the network in order to talk to another STA (fig.2).

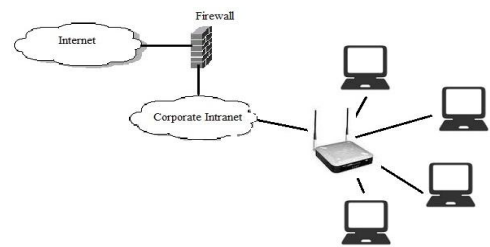

**Figure 2** Infrastructure Mode

However, in the independent mode or ad hoc mode, the STAs can directly communicate with each other. In this mode, the source STA forms an ad hoc link directly with the destination STA (fig.3).

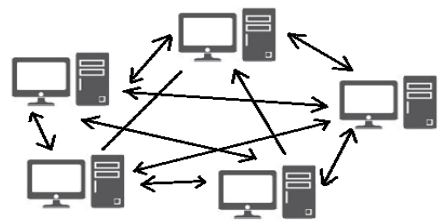

**Figure 3** Independent Mode

### 3.6. Quality of Service (QoS)

In 2005, IEEE developed a new extension of 802 standards known as 802.11e to standardize QoS enhancement efforts. This was done to tackle the QoS provisioning problem induced by the MAC techniques like PCF and DCF in the earlier extensions. The MAC techniques in the IEEE 802.11e significantly enhance QoS support for Wi-fi. Two new MAC techniques were introduced in 802.11e. These were HCF and EDCF. The HCF method was based on PCF and EDCF was based on DCF. In addition to these, two more MAC enhancements were also introduced to increase the MAC layer throughput. One of them was Block Acknowledgement which enabled sending of a single ACK for a block of frames. Another one was Direct Link Protocol, which enabled direct link between two STAs in a single WLAN network.

### 3.6.1. Hybrid Coordination Function (HCF)

The principle MAC technique in IEEE 802.11e extension, Hybrid Coordination Function is based on PCF mechanism used in the earlier standards. However unlike PCF, HCF is mandatory for devices that support 802.11e.The contention free period in HCF is known as HCF controlled Channel Access (HCCA).

In HCF, there are several EDCA periods for random access. The main difference between PCF and HCF is that in HCF unlike PCF, the packets with different QoS are mapped into different MAC queues, with different polling priorities assigned to each one. They are known as Traffic Classes (TC). The QoS enhanced Access Point (QAP) polls the TCs of the STAs using a priority based round robin schedule starting with the first QSTAs having the highest TC queue.

### 3.6.2. Enhanced Distributed Channel Access (EDCA)

EDCA is only used as a supplementary part with the first technique, HCF. It is quite similar to DCF. It is also similar in operation to HCCA as packets are queued up in different Access Categories (ACs) in EDCA similar to the TCs. In EDCA, different QSTAs act as a virtual STA for different QoS levels. The medium access parameters of ACs of a given STA depend on their respective priorities. ACs are however different from TCs; as a given type of packet always has one AC and one TC in a wi-fi network.

### 3.6.3.Block Acknowledgement

The block acknowledgement mechanism was also introduced in 802.11e standard for QoS enhancement. Traditionally, an ACK was sent for every MAC frame received by the destination STA. But ACKs caused high overhead and waiting time delay. To overcome this, IEEE 802.11e used frame blocks and sent one ACK for every block. Even though it speeds up the process but the primary disadvantage of retransmitting the entire block again on failure to receive an ACK signal still persists.

### 3.6.4. Direct Link Protocol (DLP)

Two STAs in the same WLAN can only communicate through an AP. So the amount of bandwidth consumed is twice the actual rate of transmission. DLP is a technique which allows two STAs to directly communicate with each other without using an Access Point, avoiding excess of bandwidth.

The MAC techniques in IEEE 802.11e divide the transfer time into Transmission Opportunities (TO) [10]. A QSTA can use the medium for a number of TOs. This eliminates the timing problem in PCF. In both the EDCA and HCF techniques, at the onset of transmission, the QSTA sends an information packet with QoS specifications for communication request. Now if a QAP can handle the transmission subject to all the QoS parameters passed to it, it accepts it otherwise it rejects the request, which can again be retried with a relatively lower QoS parameter.

Additional mechanisms to further improve MAC throughput and QoS support have been proposed. In the paper [13] it has been shown that using an adaptive contention window, Inter Frame Space (IFS) parameters which are based on the link condition improve the overall performance of IEEE 802.11e extension. In [12] it was proposed that the polling mechanism could be replaced with a request based scheduler. In this case, each of the QSTA sends a request for their queues to the QAP. The QSTA schedules these uplink packets with its own downlink packets using a modified Weighted Fair Queuing Scheme.

### 3.7. Contention Based Protocol (CBP)

First introduced in IEEE 802.11y the contention based protocol is a communication protocol for wireless communication equipment that allow user to use the same radio channel without pre-coordination. One of the well known contention based protocol of the IEEE 802.11 standard is "listen before talk".

The protocol allows multiple users to use the same radio frequency spectrum by establishing rules in which a transmitter provides reasonable opportunities to other transmitters when both of them simultaneously try to access the same channel. It consists of separate procedures for initiating new transmissions, determining the state of the channel and for managing retransmissions in the event of a busy channel.

### 3.8. Extended Channel Switch Announcement (ECSA)

If the necessity arises to change the current operating channel or the channel bandwidth, the IEEE 802.11y provided a mechanism for an AP to notify the stations connected to it of its intentions. This will provide an opportunity to the WLAN to choose a channel which is less noisy and the least likely to cause any interference. In addition, ECSA also provides other functionalities besides dynamic channel selection like for instance, the licensed operator can send ECSA commands to any stations operating under their control, either registered or unregistered [39]. The IEEE 802.11n also used ECSA which provided for the 20MHz and 40MHz channel switching in the 11n PHY's ECSA. However, IEEE 802.11n is specified for operation in the 2.4GHz and 5GHz license exempt bands. But after IEEE 802.11y, future amendments could permit 11n's PHY to operate in other bands as well, which is seen in the upcoming IEEE 802.11ad standard

### 3.9. Dependent Station Enablement (DSE)

Another important concept introduced in the IEEE 802.11y is the mechanism by which an operator extends and retracts permission to license exempt devices to use the licensed radio spectrum. In addition, the DSE provides benefits like enabling a dependent STA to connect to its nearest AP for a short period of time and use the internet to complete the channel permission process with the enabling station. This facility reduces the likelihood of a dependent STA to cause interference while attempting to connect to a far off enabling station.

The personal privacy and security of end users are ensured while, at the same time, licensees will have the information necessary to resolve disputes. This is achieved as the devices based on IEEE 802.11y transmit a unique identifier to resolve the interference dispute. The enabling stations also transmit their unique identifier indicating their location. The FCC database will also contain the location of the enabling station that will identify the licensee. The dependent STAs provide the location of the station that enabled it plus a unique string supplied by the enabling station. This ensures that to resolve disputes that the responsible party, the licensee, can be contacted. This method alleviates the problems associated with having the dependent STA broadcasting its location. However this method would increase the cost and complexity of devices. Moreover, this technique also suffers from the uncertainty that mobile devices that constantly beacon its location could be adversely used by third parties to track the user's location

### 3.10. Mobility

The IEEE 802.11 standard was developed as an alternative to wireline cellular networks. However the mobile user profiles were originally not defined in the IEEE 802.11 standard [18]. IEEE 802.11e addressed the issue of intra AP communication but the standard still does not provide any solution for roaming users [13]. However, the cellular IP Architecture provides a solution for seamless transition between different LAN networks [24]. Here, several WLANs are connected to the internet and are interconnected among them using a gateway, which keeps records of all the moving paths.

## 4. Other Standards

### 4.1. IEEE 802.11c

The IEEE 802.11c standard covers bridge operation. Formally, a bridge is a device which links to LANs with a similar or identical MAC protocol. A bridge performs functions which are similar to those of an Internet Protocol (IP)-level router, but at the MAC layer. A bridge is simpler and efficient than an IP router.

In 2003, the 802.11c task group completed its work on this standard, which later on folded into the IEEE 802.1d standard for LAN bridges. IEEE 802.11c acted as a supplement to IEEE 802.1d which added requirements associated with bridging 802.11 wireless client devices. It adds a sub clause under 2.5 Support of the Internal Sublayer Service, to cover bridge operations with 802.11 MACs. As of now, this standard is a part of IEEE 802.1d.

## 4.2. IEEE 802.11d

Deployed in 2001, IEEE 802.11d is an amendment to the IEEE 802.11 specification that added support for "additional regulatory domains"[8]. Issues related to regulatory differences in various countries are covered and also defined operation parameters at the 5MHz frequency band, but that only for North America, Europe and Japan. This support also included the addition of a country information element to beacons, probe requests as well as to probe responses. This country information element greatly simplified creation of 802.11 wireless APs and client devices that meet the different regulations enforced in various parts of the world to which they must conform to operate at the 5MHz.

## 4.3. IEEE 802.11e

IEEE 802.11e was deployed in 2005 and is an important amendment to the IEEE 802.11 standard that provided a set of Quality of Service enhancements for wireless LAN applications through modifications to the Media Access Control (MAC) layer [10].

As discussed earlier, IEEE 802.11e accommodated time-scheduled and polled communication during null periods when no other data is moving through the system [9]. In addition, IEEE 802.11e improves the polling efficiency and channel robustness. These enhancements should provide the quality which is necessary for services such as IP telephony and video streaming [12-14]. In a QSTA, a hybrid coordination function (HCF) replaces modules for a distributed coordination function (DCF) and point coordination function (PCF). The HCF consists of enhanced distributed-channel access (EDCA) [11] and HCF-controlled channel access (HCCA). EDCA extends the existing DCF mechanism to include priorities. As with the PCF, HCCA centrally manages medium access.

## 4.4. IEEE 802.11f

To provide for wireless access point communications among multivendor systems, the IEEE 802.11f or Inter-Access Point Protocol was a recommendation that described an optional extension to IEEE 802.11. In addition to providing communication among WLAN stations in its area, an AP can also function as a bridge that connects any two 802.11 LANs across another type of a network, such as an Ethernet, LAN or a WMAN. Thus, IEEE 802.11f facilitated the roaming of a device from one AP to another while ensuring transmission continuity. IEEE 802.11f standard was a Trial Use Recommended Practice and so the standard was later withdrawn by the IEEE 802 executive committee in 2006.

## 4.5. IEEE 802.11h

The European Union Military uses part of the 5 GHz frequency band for satellite and radar communication in addition to the IEEE 802.11a signals. So to handle such situations, IEEE 802.11h introduced two mechanisms. These were

1. Dynamic Frequency Selection (DFS)
2. Transmit Power Control (TPC)

In the DFS scheme, the AP detects other networks operating in the same frequency band and changes the operating frequency of the WLAN to prevent collision. On the hand, TPC is used to keep the signal level below a certain preset level if there is a satellite signal in the nearby channels. Moreover, TPC can also be used to improve the link condition by switching over the working frequency to a more suitable channel which is clearer and also reduce power consumption [19].

## 4.6. IEEE 802.11i

The aim of the IEEE 802.11i standard was to address security issues. The security and authentication mechanisms at the MAC layer were defined in this standard. It addressed the security deficiencies in the Wired Equivalent Privacy (WEP) algorithm originally designed for the MAC layer of 802.11 [22-23]. Wired Equivalent Privacy (WEP) was shown to have security vulnerabilities.

Wi-Fi Protected Access (WPA) was introduced by the Wi-Fi Alliance as an intermediate solution to the WEP insecurities. The Wi-Fi Alliance refers to their interoperable implementation of the full 802.11i as WPA2, also called RSN (Robust Security Network). The 802.11i standard makes use of the Advanced Encryption Standard (AES) block cipher in contrast to WEP and WPA which use the RC4 stream cipher [20].

Robust Security Network (RSN) comes with two new protocols. They are:
1. The 4-Way Handshake
2. The Group Key Handshake.

These utilize the authentication services and port access control described in the IEEE 802.1X to establish as well as change the appropriate cryptographic keys. The RSN is a security network that only allows for the creation of robust security network associations (RSNAs), a type of association used by a pair of STAs if the procedure to establish authentication or association between them includes the 4-Way Handshake. It also provides two RSNA data confidentiality and integrity protocols, TKIP and CCMP, with CCMP being mandatory [16].

### 4.7. IEEE 802.11j

802.11j standard was introduced in 2004 and was designed especially for Japanese market. It provided Wireless LAN operation in the 4.9 to 5 GHz band so as to conform to the Japanese rules for radio operation for indoor, outdoor and also mobile applications.

### 4.8. IEEE 802.11k

For the purpose of radio resource management, IEEE 802.11k defines enhancements that provide mechanisms available to protocol layers above the physical layer 23]. The standard enabled the management of the air interface between several APs, including the following:

1. To improve roaming decisions, an AP can provide a site report to a mobile device when the AP determines that the mobile device is going away from it. The site report lists APs from best to worst that a mobile device can use in changing over to another AP [24].

2. An AP can collect channel information from each mobile device on the WLAN network. Each mobile device provides a noise histogram that displays all non-802.11 energy on that channel as perceived by the mobile device. The access point collects the statistics on how long a channel is in active use during a given time. This data enable the AP to regulate access to a given channel.

3. APs can query mobile devices to collect statistics, such as retries, packets transmitted, and packets received. This gives the AP a more complete view of the network performance.

4. 802.11k also extended the transmit-power-control procedures as defined in 802.11h standard to other regulatory domains and frequency bands, to reduce power consumption, interference, and to provide range control.

### 4.9. IEEE 802.11l

This standard is reserved and will not be used. Moreover, no task group was named IEEE 802.11l to avoid confusion with another well known standard IEEE 802.11i.

### 4.10. IEEE 802.11m

An ongoing task group was put in charge with the maintenance of the standard. The revisions as well as clarifications and modifications are produced periodically. IEEE 802.11m is an ongoing initiative that provides a unified view of the 802.11 base standards through continuous monitoring, management and maintenance. IEEE 802.11m is based on two documented initiatives. They are:
1. IEEE 802.11ma
2. IEEE 802.11mb

IEEE 802.11ma is comprised of the eight previous standards (a, b, d, e, g, h, i, j) which was editorially applied through 2003. In addition, IEEE 802.11mb is the current maintenance draft version.

### 4.11. IEEE 802.11n

Finally published in 2009, the aim of the IEEE 802.11n is to increase the MAC layer throughput from the previous standards. First conceived in the year 2002, the IEEE 802.11n task group has been studying various enhancements to the physical and MAC layers to improve throughput [25]. These enhancements include such items as changes to signal encoding schemes, multiple antennas, smart antennas, and changes to MAC protocols [26-28]. IEEE 802.11n introduced new MAC layer mechanisms to increase throughput. The standard uses MIMO [26] and various new modulation and coding mechanisms to increase the data rates. The standard uses a fixed channel bandwidth of 20MHz, which is useful for backward compatibility with older standards. IEEE 802.11n also possesses an optional 40MHz channel. Data rates up to 600 Mbps are achieved only with the maximum of four spatial streams using one 40 MHz-wide channel [25].

In order to increase the MAC layer efficiency two new mechanisms were introduced in IEEE 802.11n. These were:
1. Frame Aggregation
2. Block Acknowledgement

In the previous MAC layer, a STA waits for some time after sending a MAC frame. There is severe under utilization when the MAC frames are small. The frame aggregation technique enabled STAs to aggregate small frames into larger ones. To maximize efficiency, the maximum frame size is increased thereby allowing longer frames [27]. The Block Acknowledgement method discussed earlier is similar to the mechanism with the same name as in IEEE 802.11e.

## 4.12. IEEE 802.11o

Similar to the IEEE 802.11l, no task group was named as IEEE 802.11o to avoid confusion.

## 4.13. IEEE 802.11p

To provide for wireless access in vehicular environments (WAVE), IEEE 802.11p is an approved amendment to the IEEE 802.11 standard. It defines enhancements to main 802.11 required to support Intelligent Transportation Systems (ITS) applications [29-31]. It includes data exchange between high-speed passenger vehicles and ambulances and the roadside infrastructure in the licensed ITS band of 5.9 GHz (5.85-5.925 GHz).In this protocol, the vehicles send information about their traffic parameters like speed, distance from other vehicles etc. to nearby vehicles. In this way each vehicle in the corresponding neighborhood knows about the traffic status and acts accordingly.

## 4.14. IEEE 802.11r

Published in the year 2008, IEEE 802.11r or fast BSS transition (FT) enabled wireless connectivity with secure and fast handoffs between STAs [32]. This standard provided fast roaming, even for vehicles in motion. It reduced the roaming delay between two basic service sets (BSS) to less than 50 ms. The standard also refined the mobile client transition process between APs by redefining the security key negotiation protocol, which permit negotiations and requests for wireless resources. IEEE 802.11r key strength lies in the IEEE 802.1X security support, which facilitates the deployment of portable phones with VoIP over Wi-Fi.

## 4.15. IEEE 802.11s

IEEE 802.11s is an IEEE 802.11 standard for mesh networking. It defines how wireless devices can interconnect to create a WLAN mesh network, which may be used for static topologies as well as for ad-hoc networks [36]. Initially in 2003, the 802.11s started as a Study Group of IEEE 802.11. It became a Task Group in 2004. In the following year i.e. in 2005, a call for proposals was issued, which resulted in the submission of 15 proposals submitted to a vote. After a series of eliminations and mergers, the proposals dwindled to two (the "SEE-Mesh" and "Wi-Mesh" proposals), which became a joint proposal in 2006 (fig.4).

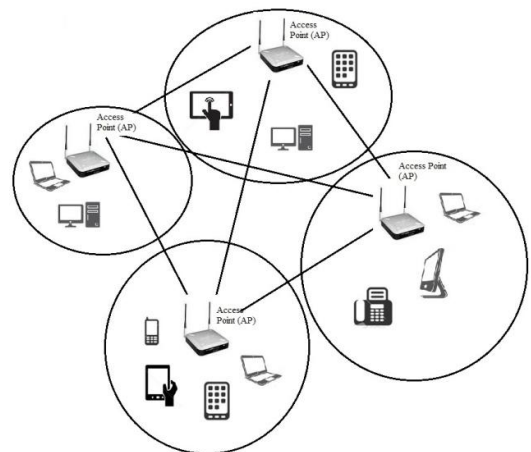

**Figure 4** Mesh Mode Configuration

Network device which is based on the IEEE 802.11s is labeled as mesh STA. Mesh STAs form mesh links with each other, over which mesh paths can be established using a routing protocol (fig.4). The 802.11s defines a default mandatory routing protocol Hybrid Wireless Mesh Protocol or HWMP [37]. But it still allows vendors to operate using alternate protocols. HWMP is inspired from a combination of AODV and tree-based routing.

Mesh STAs being individual devices use mesh services to communicate with other devices in the WLAN network. They can also collocate with 802.11 APs and provide access to the mesh network. Also, mesh STAs can collocate with an 802.11 portal that implements the role of a gateway and provides access to one or more non-802.11 networks. In both the cases, 802.11s provides a proxy mechanism to provide addressing support for non-mesh 802 devices, allowing for end-points to be cognizant of the external addresses.

802.11s also provides mechanisms for deterministic network access, a framework for congestion control and power save [34-36]. There are no defined roles in a mesh — no clients and servers, no initiators and also no responders. Security protocols used in a mesh must, therefore, must be true peer-to-peer protocols where either side can initiate to the other or both sides can initiate simultaneously.

A key establishment protocol called "Simultaneous Authentication of Equals" (SAE) defines a secure password-based authentication technique. SAE is based on Diffie–Hellman key exchange using finite cyclic groups which can be a primary cyclic group or an elliptic curve. The problem on using Diffie–Hellman key exchange is that it does not have an authentication mechanism. Hence to solve the authentication problem, the resulting key is influenced by a pre-shared key as well as the MAC addresses of both peers [34].

SAE exchange takes place only when both the peers discover each other. On successful completion of SAE, each peer knows that the other party possesses the mesh password and, as a by-product of SAE exchange, the two peers establish a cryptographically strong key. The key is then used with the "Authenticated Mesh Peering Exchange" (AMPE) to establish a secure peering and derive a session key to protect mesh traffic, including routing traffic [35].

### 4.16. IEEE 802.11t

The IEEE 802.11t standard regroups recommended practices to test and measure performance in wireless networks. It was also known as WPP (Wireless Performance Prediction). However as on date, the standard has been cancelled from the IEEE 802 project.

### 4.17. IEEE 802.11u

Published in the year 2011 the IEEE 802.11u was developed to improve internetworking with external non-802.11 networks. This was achieved through network discovery and selection and QoS map distribution. The network discovery and selection may be provided in the following ways:

1.  The discovery of suitable networks through the advertisement of access network type, roaming consortium and venue information.
2.  Generic Advertisement Service (GAS) provides for Layer 2 transport of an advertisement protocol's frames between a mobile device and a server in the network prior to authentication. The AP is responsible for the relay of a mobile device's query to a server in the carrier's network and for delivering the server's response back to the mobile.
3.  Access Network Query Protocol (ANQP), which is a query and response based protocol used by a mobile device to discover a range of information, like roaming partners accessible via the hotspot along with their credential type including the hotspot operator's domain name; and EAP method supported for authentication; IP address type availability and other metadata useful in a mobile device's network selection process.

The QoS provides a mapping between the IP's differentiated services code point (DSCP) to over-the-air Layer 2 priority on a per-device basis, facilitating end-to-end QoS [38]. Nowadays, with mobile users whose devices can move between 3G and Wi-Fi networks at a low level using 802.21 handoff, there arise needs for a unified and reliable way to authorize their access to all of those networks. The 802.11u standard provides a common abstraction that all networks regardless of protocol can use to provide a common authentication experience. Actually, IEEE 802.11u primarily focuses on on-the-fly authorization between both the STA and the AP [38]. With the usage of external network authorization, the AP also provides service to the previously unknown STAs.

### 4.18. IEEE 802.11v

The IEEE 802.11v was introduced in 2011, to enable configuring clients while they are connected to the network.

### 4.19. IEEE 802.11w

The IEEE 802.11w was published in 2009 as an add-on to 802.11i covering management frame security. It introduces protected management frames with the help of mechanisms that enable data origin authenticity, data integrity and replay protection.

### 4.20. IEEE 802.11x

This standard was also not used to avoid confusion with IEEE 802.1x.

### 4.21. IEEE 802.11y

Since the frequency channels at the 2.4 and 5 GHz frequency bands are already occupied by the WLANs, a new frequency band was defined in the US allowing higher power and thus longer ranges [39]. The IEEE 802.11y adapts current layers to the new frequency band. To allow for this, the standard also adds three new concepts:

1. Contention Based Protocol (CBP)
2. Extended Channel Switch Announcement (ECSA)
3. Dependent Station Enablement (DSE)

These have already been discussed in section 3 and explained in details in [39-40].

### 4.22. IEEE 802.11z

Published in 2010, the IEEE 802.11z provides extensions for Direct Link Setup (DLS). DLS allows two stations to communicate directly with each other without the requirement of an AP. This standard basically provides direct-link setup enhancements to the IEEE 802.11 MAC and PHY, extending direct-link setup to be independent of the AP, in addition to adding power save capabilities. The direct-link setup is made independent of the AP by tunneling the protocol messages inside data frames.

### 4.23. IEEE 802.11aa

The IEEE 802.11aa specifies enhancements to the IEEE 802.11 medium access control (MAC) for robust audio video (AV) streaming, at the same time maintaining coexistence with other types of traffic.

## 5. Upcoming Standards

### 5.1. IEEE 802.11ad

The IEEE 802.11ad also known as WiGig is a relatively new standard published in December 2012. It specification adds a "fast session transfer" feature, enabling the wireless devices to seamlessly transition between the legacy 2.4 GHz and 5 GHz bands and the 60 GHz frequency band [41]. To operate with optimal performance and range criteria, the IEEE 802.11ad provides the ability to move between the bands ensuring that computing devices are always "best connected,"

Through the vast improvements in spectral reuse at 60 GHz and an efficient beam forming technology, IEEE 802.11ad enables great improvements in capacity [42]. Many users in a dense deployment can all maintain top-speed performance, without interfering with each other or having to share bandwidth as with the legacy frequency bands [43-44]. The likely enhancements to 802.11 beyond a new 60 GHz PHY include MAC modifications for directional antennas, personal basic service set, beamforming, fast session transfer between PHYs and spatial reuse.

### 5.2. IEEE 802.11ae

The IEEE 802.11ae introduces a mechanism for prioritization of management frames. A protocol to communicate management frame prioritization policy is specified in this standard.

### 5.3. IEEE 802.11ac

One of the important standards currently under development is IEEE 802.11ac. This standard is expected to be published by the end of 2014. It is expected to provide a multi-station WLAN throughput of at least 1 Gbps and a single link throughput of at least 500 Mbps [45]. This is achieved by extending the air interface concepts which are embraced by 802.11n like wider RF bandwidth (up to 160 MHz), more MIMO spatial streams (up to 8), multi-user MIMO, and high-density modulation [45-47].

### 5.4. IEEE 802.11af

The IEEE 802.11af is intended to operate in the TV White Spaces, which is the spectrum already allocated to the TV broadcasters but not in use at a specific location and time period [12]. It will use cognitive radio technology to identify white spaces it can use. However, this cognitive technology will be based on an authorized geolocation database. This database will provide information on which frequency, at what time and under what conditions networks may operate.

## 5.5. IEEE 802.11ah

The IEEE 802.11ah is aimed at developing a global WLAN network that will allow user to access sub carrier frequencies below 1GHz in the ISM band. It will also enable devices based on the IEEE 802.11 standards to get access to short burst data transmissions like meter data. In addition it will provide improve coverage range which will allow new applications such as wide area based sensor networks, sensor backhaul systems and potential Wi-Fi off-loading functions to emerge [48]. This standard is currently under development and is predicted to be finalized by 2016.

## 5.6. IEEE 802.11ai

The IEEE 802.11ai is an upcoming standard predicted to be finalized by 2015. It will include a fast initial link setup (FILS) that could enable an STA to achieve a secure link setup which is less than 100 ms [49]. A successful link setup process will then allow the STA to send IP traffic with a valid IP address through the AP.

## 5.7. IEEE 802.11mc

The IEEE 802.11mc is similar to the IEEE 802.11m and is also scheduled to appoint a working group with the task of maintenance of the standard around 2015.

## 5.8. IEEE 802.11aj

The IEEE 802.11aj will provide modifications to the IEEE 802.11ad Physical (PHY) layer and the Medium Access Control (MAC) layer to enable operation in the Chinese Milli-Meter Wave (CMMW) frequency bands including the 59-64 GHz frequency band. The amendment is also intended to maintain backward compatibility with 802.11ad when it operates in the 59-64 GHz frequency band. The standard shall also define modifications to the PHY and MAC layers to enable the operation in the Chinese 45 GHz frequency band. This standard is scheduled to be finalized by the end of 2016.

## 5.9. IEEE 802.11aq

The WLAN is fast evolving and is no longer one where stations are merely looking for only access to internet service. This creates opportunities to deliver new services, as the IEEE 802.11 standard needs to be enhanced to better advertise and describe these new services.
The IEEE 802.11aq will provide mechanisms that assist in pre-association discovery of services by addressing the means to advertise their existence and enable delivery of information that describes them. This information about services is to be made available prior to association by stations operating on IEEE 802.11 wireless networks. This standard is scheduled to be published by 2015.

## 6. Problems with IEEE 802.11

Several extensions of the original IEEE 802.11 standard had been developed but still there are many problems associated with the standards which are required to be addressed.

## 6.1. Security

One of the prime concerns in wireless networking is security. As WLANs operate over the shared medium, eavesdropping by unauthorized people and critical information may be accessed with the use of malicious technologies. The initial standard WEP had security flaws which lead the wi-fi forum to implement another encryption system WPA and later WPA2. Although WPA and WPA2 are much more secure and provides good protection still it is not secure enough to be contend with. More complex encryption algorithms need to be implemented without decreasing the MAC layer throughput.

## 6.2. Data Rate

Another serious drawback of IEEE 802.11 wireless network is the data rate, which is quite low as compared to its wireline IEEE 802.3 counterpart. IEEE 802.11g provides a data rate of 54Mbps. There was significant improvement in the data rates provided by the IEEE 802.11n extension. The WWise and TGnSync proposal supported data rates upto 540Mbps and 630Mbps respectively. However, the 40 MHz channel required to support such data rates are not available in many countries. The more recent IEEE 802.11ac aims at providing data rates upto 1Gbps. But this data rate is even still quite low as compared to the 10 GHz capacity of the wireline network standard 802.3n.

## 6.3. MAC Layer Throughput

Even though the IEEE 802.11g provides a data rate of 54 Mbps, the MAC layer throughput is far from this value. Detailed studies have shown that with higher data rates, the MAC layer throughput decreases to 40-50% of the new data rate. A number of factors can be held responsible for this decrease in throughput such as the link condition overhead of the MAC layer control headers and usage of the RTS/CTS mechanism. Increasing the PHY layer data rate alone cannot solve the throughput problem; but the MAC layer also needs to be changed. Two extensions IEEE 802.11e and IEEE 802.11n both introduce new mechanisms to solve the problem, but these techniques are far from providing a complete solution.

## 7. ETSI HiperLAN family

Part of the ETSIs BRAN project, is the High Performance Radio Local Area network (HiperLAN) which is also targeted as wireless local area network. HiperLAN/1 the first standard to be developed in 1996 supported data rates upto 20Mbps. In 2000, another standard HiperLAN/2 was deployed which provided data rates upto 54 Mbps. The HiperLAN networks were developed with more detailed MAC layer mechanisms than it IEEE counterpart.

HiperLAN did not enjoy much success as the IEEE standard in spite of offering high data rates initially than some of the earlier extensions like IEEE 802.11a and improved MAC layer mechanisms [50]. No new HiperLAN standards are currently reported to be under development

## 7.1. Physical (PHY) Layer

HiperLAN/1 was designed to operate at a frequency band of 5.15-5.35 GHz. Similarly HiperLAN/2 also operated at a frequency band of 5.47-5.725 GHz [52]. Unlike IEEE 802.11, both the HiperLAN standards had a fixed common channel bandwidth of 20MHz. HiperLAN 802.11a had a similar number of non-overlapping channels. A HiperLAN network chooses its own channel by using a DFS mechanism, which changes the operating frequency of the network when the current frequency is occupied by another network. The HiperLAN PHY layer uses OFDM and several modulation techniques along with coding values for different data rates. Table 2 summarizes the various modulation techniques of the PHY layer.

The range of operation of HiperLAN is 30m indoors to 150m outdoors. Use of omnidirectional and sectorized antennas could be seen in a HiperLAN BS or STA.

**Table 2**
HiperLAN PHY layer modulation techniques

| Data Rate (Mbps) | Modulation | Coding rate | Coded bits/sub carrier |
|---|---|---|---|
| 6 | BPSK | 1/2 | 1 |
| 9 | BPSK | 3/4 | 1 |
| 12 | QPSK | 1/2 | 2 |
| 18 | QPSK | 3/4 | 2 |
| 27 | 16-QAM | 9/16 | 4 |
| 36 | 16-QAM | 3/4 | 4 |
| 54 | 64-QAM | 3/4 | 6 |

## 7.2. Data Link Control (DLC) Layer

Similar to its IEEE counterpart, there are two operating modes in HiperLAN also. These are:
1. Centralized Mode
2. Direct Mode

In the centralized mode, all of the network traffic is routed through an AP. In the direct mode, on the other hand if both the source and the destination for a transmission are in the same network, the connection is made directly between the source and the destination node. Like in the centralized mode, the direct mode employs a Central Controller (CC) which is responsible for the management of all the traffic in the network [50].

DLC layer of HiperLAN uses a Time Division Multiple Access Scheme (TDMA). In HiperLAN each MAC frame is divided into five phases. They are:
1. Broadcast Phase
2. Downlink Phase
3. Direct Link Phase
4. Uplink Phase
5. Random Access Phase

The first phase which is the broadcast phase mainly contains the broadcast messages of the network. In addition it also contains the frame structure information i.e. the time information of both the uplink and downlink phase of the current frame. The uplink and the downlink phases contain messages which originate from the AP/CC as well as from a WLAN node. If a connection is specified to be a direct link connection, frames belonging to this connection are sent in the Direct Link Phase. Association messages and resource allocation messages is transmitted in the Random Access Phase.

In HiperLAN when a STA needs to establish a connection a request for a new connection is sent to the corresponding AP/CC. IF the AP/CC accepts this request based on the QoS constraints of the connection and the current network load, it establishes the connection and reserves it resources accordingly. Otherwise the request is denied. In case of a denial, the STA may retry with a new request. The QoS connections are set up during connection establishment. There is no MAC layer throughput limit in HiperLAN.

HiperLAN uses a plane based protocol stack. In this, two planes exist in the protocol where one is for user data and the other is for control messages. All the data packets are sent over the user plane and the control packet over the control plane.

## 7.3. Security

. The security in HiperLAN/1was covered in the MAC layer. However HiperLAN/2 addressed the issue of security with emphasis. The HiperLAN/2 had security functions in the DLC level for protection against eavesdropping and transmission inception. This was established with the help of a Data Confidentiality service in the DLC layer. It deployed two encryption-decryption methods, DES and later triple DES. However these techniques proved to be quite insecure over time.

## 8. HiperLAN Standards

## 8.1. HiperLAN/1

The goal of HiperLAN/l was to achieve higher data rates than the ongoing IEEE standard The HiperLAN/1 employs the same Physical layer and the Media Access Control part of the Data link layer like 802.11. In addition, HiperLAN provides a new sub-layer called Channel Access and Control sublayer (CAC). This sub-layer deals with the access requests which are made to the networks. The accomplishing of the request is dependent on the usage of the network and the priority of the request.

CAC layer provides hierarchical independence with Elimination-Yield Non-Preemptive Multiple Access mechanism (EY-NPMA) [51]. EY-NPMA codes priority choices and other functions into one variable length radio pulse preceding the packet data, enabling the network to function with few collisions even with a large number of users. Multimedia applications work in HiperLAN because of EY-NPMA priority mechanism. The MAC layer defines protocols for security, routing and power saving and provides for naturally data transfer to the upper layers. On the physical layer FSK and GMSK modulations are used in HiperLAN/1.

## 8.2. HiperLAN/2

HiperLAN/2 functional specification was accomplished in 2000. The physical layer of HiperLAN/2 is very similar to IEEE 802.11a standard. However, the media access control is Dynamic TDMA in HiperLAN/2, while CSMA/CA is used in 802.11a [53].The standard covers Physical, Data Link Control and Convergence layers.

The Convergence layer takes care of service dependent functionality between DLC and Network layer. Convergence sub-layers can be used also on the physical layer to connect IP, ATM or UMTS networks [52]. This feature makes HiperLAN/2 suitable for the wireless connection of various networks. HiperLAN/2 physical layer makes use of BPSK, QPSK, 16QAM or 64QAM modulations. HiperLAN/2 for the first time offered security measures. The data were secured with DES or Triple DES algorithms.

## 9. Failure in the Market

HiperLAN was basically a European alternative for the IEEE standard. Due to competition from IEEE 802.11, this was simpler to implement and was released prior to HiperLAN [50]. HiperLAN never received much commercial

**Table 3**

IEEE 802.11 Family

| Standard | Purpose | Publishing Date |
|---|---|---|
| IEEE 802.11 | Originally 1 Mbps and 2 Mbps, 2.4 GHz RF and IR standard | 1997 |
| IEEE 802.11a | 54 Mbps, 5 GHz PHY layer standard | 1999 |
| IEEE 802.11b | Enhancements to 802.11 to support 5.5 and 11 Mbps | 1999 |
| IEEE 802.11c | Bridge operation procedures [**now included in the IEEE 802.1D**] | 2001 |
| IEEE 802.11d | Country-to-country roaming extensions | 2001 |
| IEEE 802.11e | Enhancements: QoS, including packet bursting | 2005 |
| IEEE 802.11f | Inter-Access Point Protocol [**Stands Cancelled**] | 2003 |
| IEEE 802.11g | 54 Mbps, 2.4 GHz standard (backwards compatible with b) | 2003 |
| IEEE 802.11h | Spectrum Managed 802.11a (5 GHz) for European compatibility | 2004 |
| IEEE 802.11i | Enhanced security | 2004 |
| IEEE 802.11j | Extensions for Japan | 2004 |
| IEEE 802.11k | Radio resource measurement enhancements | 2007 |
| IEEE 802.11m | Maintenance of the standard | |
| IEEE 802.11n | Higher throughput improvements using MIMO | 2009 |
| IEEE 802.11p | WAVE—Wireless Access for the Vehicular Environment | 2010 |
| IEEE 802.11r | Fast BSS transition (FT) | 2008 |
| IEEE 802.11s | Mesh Networking, Extended Service Set (ESS) | 2011 |
| IEEE 802.11t | Wireless Performance Prediction (WPP)—test methods and metrics Recommendation [**Stands Cancelled**] | |
| IEEE 802.11u | Improvements related to Hot Spots and 3rd party authorization of clients. | 2011 |
| IEEE 802.11v | Wireless network management | 2011 |
| IEEE 802.11w | Protected Management Frames | 2009 |
| IEEE 802.11y | 3650–3700 MHz Operation in the U.S. | 2008 |
| IEEE 802.11z | Extensions to Direct Link Setup | 2010 |
| IEEE 802.11aa | Robust streaming of Audio Video Transport Streams | 2012 |
| IEEE 802.11ad | Very High Throughput 60 GHz | 2012 |
| IEEE 802.11ae | Prioritization of Management Frames | 2012 |
| IEEE 802.11ac | Very High Throughput <6 GHz; potential improvements over 802.11n: better modulation scheme (expected ~10% throughput increase), wider channels, multi user MIMO. | ~2014 |
| IEEE 802.11af | TV Whitespace | ~2014 |
| IEEE 802.11ah | Sub 1 GHz sensor network, smart metering. | ~2016 |
| IEEE 802.11ai | Fast Initial Link Setup | ~2015 |
| IEEE 802.11mc | Maintenance of the standard | ~2015 |
| IEEE 802.11aj | China Millimeter Wave | ~2016 |
| IEEE 802.11aq | Pre-association Discovery | ~2015 |

~ denotes expected year

implementation. Much of the work on HiperLAN/2 has lasted in the PHY specification for IEEE 802.11a, which is nearly identical to the PHY specification of HiperLAN/2.

## 10. Conclusion

Some of the reasons which can be cited for such widespread use of WLANs are low infrastructure cost, ease of development, support for mobile user communication, deployment without cabling and ease of adding new user to the network resulting in a huge decrease in implementation cost. As the importance of mobile user has increased manifold, WLANs have gained much importance in homes, schools, offices etc and matured as an access technology in short distance communications. Still today, WLANs suffers from a lot of problems. One of the most important drawbacks is the use of shared medium in which performance gets considerably degraded as the number of STAs increases in the WLAN network. The problem of unauthorized access and eavesdropping in WLANs are some of the serious security issues which have been a long standing headache for the IEEE working group. Different security encryption schemes had been implemented in the past. But so far all such encryption systems have been proven to have security vulnerabilities.

Furthermore a WLAN network provides a contention based access to its users. Contention based access is a probabilistic access approach where each user in the network competes with others for access through an Access Point (AP). This approach becomes cumbersome even with a small network load as small as a classroom. Maintaining a strict QoS is quite troublesome due to contention and variable link quality problems. Various methods have been implemented to overcome these problems and more advanced methods are currently under development. Brief overviews of the standards that belong to the IEEE 802.11 have been summarized in Table 3.

## Abbreviations

| Abbreviations | Open form |
| --- | --- |
| AC | Access Category |
| ACK | Acknowledgement |
| AES | Advanced Encryption Standard |
| ANQP | Access Network Query Protocol |
| AODV | Ad hoc On-Demand Distance Vector |
| AP | Access Point |
| ARQ | Automatic Response Request |
| ATM | Autonomous Transfer Mode |
| BPSK | Binary Phase Shift Keying |
| BRAN | Broadband Radio Access Network |
| BSS | Basic Service Sets |
| CAC | Channel Access and Control sublayer |
| CBP | Contention Based Protocol |
| CC | Central Controller |
| CCMP | Counter Cipher Mode with Block Chaining Message Authentication Code Protocol |
| CD | Collision Detection |
| CSMA | Carrier Sense Multiple Access |
| DFS | Dynamic Frequency Selection |
| DLP | Direct Link Protocol |
| DLC | Data Link Control |
| DLS | Direct link Setup |
| DSCP | Differentiated Services Code Point |
| DSE | Dependent Station Enablement |
| DSSS | Direct Sequence Spread Spectrum |
| ECSA | Extended Channel Switch Announcement |
| EDCA | Enhanced Distributed Coordination Access |
| ETSI | European Telecommunications Standards Institute |
| EY-NPMA | Elimination-Yield Non-Preemptive Multiple Access mechanism |
| FHSS | Frequency Hopping Spread Spectrum |
| GAS | Generic Advertisement Service |
| HCCA | HCF controlled Channel Access |
| HCF | Hybrid Coordination Function |
| HiperLAN | High performance Radio Local Area Network |
| HWMP | Hybrid Wireless Mesh Protocol |
| IFS | Inter Frame Space |
| IR | Infrared |
| MAC | Medium Access Control |
| MIMO | Multi Input Multi Output |
| NAK | Negative Acknowledgement |
| OFDM | Orthogonal Frequency Division Multiplexing |
| P2P | Peer to Peer |
| PCF | Point Coordination Function |
| QAP | QoS enhanced Access Point |
| QoS | Quality of Service |
| QSTA | QoS enhanced Station |
| RSN | Robust Security Network |
| RSNA | Robust Security Network Association |
| STA | Station |
| SAE | Simultaneous Authentication of Equals |
| TC | Traffic Classes |
| TDMA | Time Division Multiple Access |
| TG | Task Group (TGw) |
| TKIP | Temporal Key Integrity Protocol |
| TO | Transmission Opportunities |
| TPC | Transmit Power Control |
| UMTS | Universal Mobile Telecommunication System |
| WEP | Wired Equivalent Piracy |
| Wi-Fi | Wireless Fidelity |
| WPA | Wi-Fi protected Access |
| WPP | Wireless Performance Prediction |
| WLAN | Wireless Local Area Network |
| WWAN | Wireless Wide Area Network |